# Mathematical Modeling, In-Human Evaluation and Analysis of Volume Kinetics and Kidney Function after Burn Injury and Resuscitation

Ghazal ArabiDarrehDor, Ali Tivay, Chris Meador, George C. Kramer, Jin-Oh Hahn, *Senior Member, IEEE* [*], and Jose Salinas

G. ArabiDarrehDor, A. Tivay, and J.O. Hahn[*] are with the Department of Mechanical Engineering, University of Maryland, College Park, MD, USA (correspondence e-mail: jhahn12@umd.edu). C. Meador is with Arcos, Inc., Missouri City, TX, USA., G.C. Kramer is with the Department of Anesthesiology, University of Texas Medical Branch, Galveston, TX, USA., J. Salinas is with U.S. Army Institute of Surgical Research, San Antonio, TX, USA.

*Abstract*— **Objective**: Existing burn resuscitation protocols exhibit a large variability in treatment efficacy. Hence, they must be further optimized based on comprehensive knowledge of burn pathophysiology. A physics-based mathematical model that can replicate physiological responses in diverse burn patients can serve as an attractive basis to perform non-clinical testing of burn resuscitation protocols and to expand knowledge on burn pathophysiology. We intend to develop, optimize, validate, and analyze a mathematical model to replicate physiological responses in burn patients. **Methods**: Using clinical datasets collected from 233 burn patients receiving burn resuscitation, we developed and validated a mathematical model applicable to computer-aided in-human burn resuscitation trial and knowledge expansion. Using the validated mathematical model, we examined possible physiological mechanisms responsible for the cohort-dependent differences in burn pathophysiology between younger versus older patients, female versus male patients, and patients with versus without inhalational injury. **Results**: We demonstrated that the mathematical model could replicate physiological responses in burn patients associated with wide demographic characteristics and injury severity, and that an increased inflammatory response to injury may be a key contributing factor in increasing the mortality risk of older patients and patients with inhalation injury via an increase in the fluid retention. **Conclusion**: We developed and validated a physiologically plausible mathematical model of volume kinetic and kidney function after burn injury and resuscitation suited to in-human application. **Significance**: The mathematical model may provide an attractive platform to conduct non-clinical testing of burn resuscitation protocols and test new hypotheses on burn pathophysiology.

# INTRODUCTION

BURN is a leading cause of unintentional injury and death in the United States. According to a recent fact sheet from the American Burn Association, burn injury is the 8th leading cause of death in adults >65 years old, 3rd in children 5-9 years old, and 5th in children 1-4 years old [1]. Each year, there are nearly 500,000 burn injury incidences only across the United States, among which 40,000 are hospitalized. Burn injury results in severe inflammatory responses that lead to increased leakage of intravascular water into the tissues, which, if not treated, engenders fatal consequences such as hypovolemic shock, ischemia, multiple organ failure, and generalized edema [2], [3]. The majority of burn patients treated in burn centers survive, but many of them suffer from complications [4], [5].

Currently, treatment of burn patients involves a resuscitation protocol in which the dose of the resuscitation fluid is titrated frequently to maintain the urinary output (UO) response in a burn patient to a therapeutic target range (e.g., 30-50 ml/hr or 0.5-1.0 ml/kg·h [6]) [7]. Yet, existing resuscitation protocols exhibit a large variability in treatment efficacy, due to many factors such as the inter-patient differences in response to burn injury and resuscitation as well as the incomplete knowledge of the pathophysiology underlying burn injury, which collectively complicates therapeutic decision-making. Hence, there is an ongoing effort to optimize burn resuscitation protocols [8]–[10]. However, due to the ongoing limitations, today's burn resuscitation often starts with an established burn resuscitation protocol but is subsequently titrated in an ad-hoc fashion to the physiological responses (including UO) in an individual patient to hopefully optimize the therapeutic efficacy. Unfortunately, current treatments tend to over-resuscitate the patients [11], which would exacerbate edema and expose the patients to an elevated risk of side effects, e.g., pulmonary edema, limb and abdominal compartment syndrome, necrosis, and death, due to the accumulation of resuscitation fluid (known as "fluid creep") [12], [13]. Hence, burn resuscitation regimens must be optimized based on complete knowledge of burn pathophysiology in order to best maintain organ functions in burn patients while minimizing adverse complications.

Development and optimization of burn resuscitation protocols is challenging, because (i) burn injury is less common than other widespread injuries (i.e., protocol optimization is hampered by the small number of burn injury cases) and (ii) it is unethical to test a new treatment protocol with unproven efficacy and safety in critically ill burn patients. From this standpoint, a mathematical model that can replicate physiological responses to burn injury and resuscitation in diverse burn patients can serve as an attractive basis to (i) perform non-clinical testing of emerging burn resuscitation protocols as medical digital twin [14], [15] and (ii) expand knowledge on burn pathophysiology [16]. However, existing mathematical models are not yet ideal for such purposes for at least three reasons. First, some mathematical models cannot be applied to test many existing burn resuscitation protocols based on UO, simply because they were not developed to predict UO responses to burn injury and resuscitation [17]–[21]. Second, most existing mathematical models have not been extensively proven in humans. In fact, the ability of existing mathematical models to replicate UO and other physiological responses to burn injury and resuscitation was validated in a prohibitively small number of patients [22]–[27] or only at the group level [17]–[21]. Third, some mathematical models (especially those reported early) [22]–[27] do not reflect up-to-date knowledge of burn-related physiology and pathophysiology gained in recent experimental and clinical investigations, regarding in particular the burn-induced perturbations in volume kinetics, kidney function, lymphatic flow, and tissue pressure-volume relationships [2], [28]–[30]. Hence, closing these gaps may lead us to an enhanced mathematical model of burn injury and resuscitation ideally suited to the development and testing of emerging burn resuscitation protocols and algorithms as well as to the expansion of our knowledge of burn pathophysiology.

In this paper, we intend to develop, extensively validate, and analyze a mathematical model capable of replicating volume kinetic and kidney function responses to burn injury and resuscitation in burn patients. By leveraging clinical datasets collected from 233 real burn patients receiving resuscitation, we developed a mathematical model suited to computer-aided in-human burn resuscitation trial and knowledge expansion, by expanding our prior work and utilizing systematic parametric sensitivity analysis and regularization. We investigated the validity of the mathematical model by testing its physiological plausibility in a dedicated test dataset. Using the validated mathematical model, we examined possible mechanisms responsible for the cohort-dependent differences in burn pathophysiology by comparing the mathematical models fitted exclusively to younger versus older patients, female versus male patients, and patients with versus without inhalational injury. To the best of our knowledge, our mathematical model may be the first mathematical model extensively validated for use as digital twin of real burn patients.

This paper is organized as follows. Section II provides an overview of the mathematical model as well as the details of the clinical dataset and data analysis. Section III presents the results. Section IV is devoted to the discussion of major findings. Section V concludes the paper with future directions.

# Materials and Methods

## Mathematical Model

We continued to develop a mathematical model capable of predicting volume kinetic and kidney function responses to burn injury and resuscitation developed in our prior work [31]. The mathematical model consists of (i) volume kinetics to replicate water volume and protein concentration dynamics in the intravascular and the tissue compartments ("Volume Kinetics" in Fig. 1(a); see Appendix A), (ii) kidney functions to replicate UO response to changes in intravascular water and protein volumes ("Kidneys" in Fig. 1(a); see Appendix B), and (iii) transient perturbations in volume kinetics induced by burn as a chain of biochemical, molecular, and mechanical events ("Burn Perturbations to Volume Kinetics" in Fig. 1(b); see Appendix C).

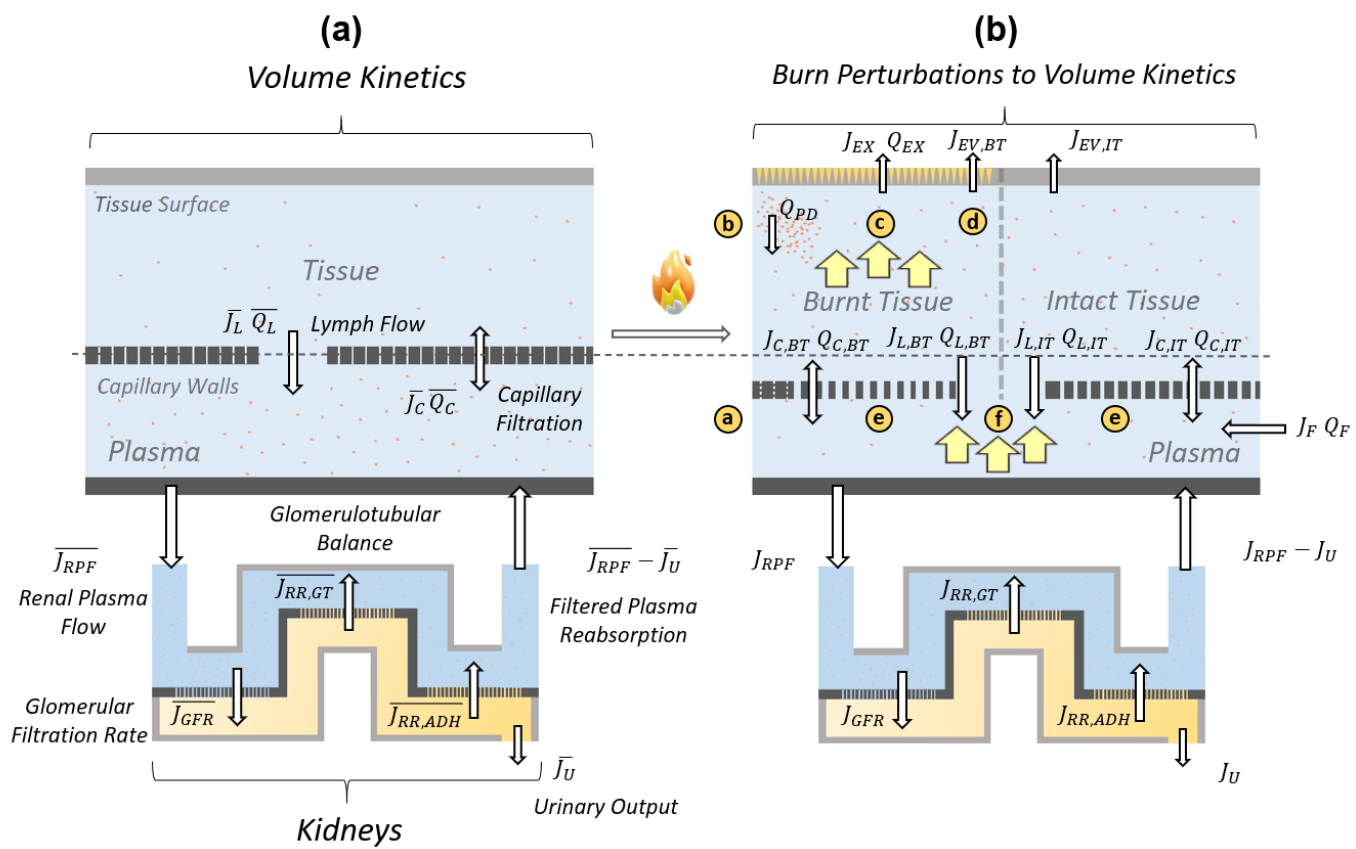

Fig. 1. Mathematical model capable of predicting volume kinetic and kidney function responses to burn injury and resuscitation. Fig. 1(a) shows normal state, where the water (blue background) and protein (pink dots) balance between the intravascular compartment ("Plasma") and the tissue is preserved by the capillary filtration through the capillary pores (thin rectangles in the capillary wall) and lymph flow. The plasma is filtered by the renal plasma flow into the kidney by way of the renal regulatory mechanisms including the glomerulotubular balance and the Antidiuretic hormone. Upon the onset of burn injury, the water and protein volumes in the plasma decrease, while tissue compartment is divided into intact and burnt tissue compartments associated with water and protein content higher than normal state ("Burnt Tissue" and "Intact Tissue") (Fig.1(b)). J: water flow. Q: albumin flow. Subscripts: C (capillary filtration); L (lymph flow); F (fluid infusion); U (UO); RPF (renal plasma flow); GFR (glomerular filtration rate); RR,GT (reabsorption rate by glomerulotubular balance); RR,ADH (reabsorption rate by Antidiuretic hormone); EX (exudation); EV (evaporation); PD (protein denaturation); BT (burnt tissues); IT (intact tissues). The yellow circles represent the perturbations that lead to the redistribution of water and protein. a: Partial destruction of capillaries in burnt tissues (shown as occluded capillary pores). b: Denaturation of protein in burnt tissues. c Transient negative hydrostatic pressure in burnt tissues, which draws water into the burnt tissues. d: Increased dermal fluid loss. e: Time-varying changes in capillary filtration and albumin permeability (shown as enlarged capillary pores). [f]: Vasodilation, which pushes water out of plasma. "X" ¯ represents "X" in normal (pre-burn) state.

The volume kinetics was represented using a three-compartmental model with intravascular space (including arterial and venous vessels, shown as "Plasma" in Fig. 1), intact tissues, and burnt tissues as separate compartments. These compartments describe the volumes of water ($V_P, V_{IT}, V_{BT}$; whose dynamics is governed by Eq. (4)) and protein therein ($A_P, A_{IT}, A_{BT}$; whose dynamics is governed by Eq. (5)) as functions of time. It was assumed that albumin serves as the surrogate of all the protein contributing to capillary filtration and colloid oncotic balance [32]. As shown in Fig. 1, the change in these volumes are dictated by (i) inter-compartmental flows of the lymphatic drainage (water $J_L$ and protein $Q_L$; Eq. (8)-(9)) and capillary filtration ($J_C$ and $Q_C$; Eq. (6) and Eq. (10)); (ii) external inputs representing the gain via burn resuscitation ($J_F$ and $Q_F$); and (iii) outputs representing the kidneys' net filtration and reabsorption of renal plasma flow (i.e., UO $J_U$; Eq. (15) and Eq. (26)), as well as burn-induced evaporation ($J_{EV}$; Eq. (34)) and exudation ($J_{EX}$ and $Q_{EX}$; Eq. (35)). The kidney function was represented by a lumped-parameter model developed in our prior work [31], which includes a hybrid combination of first-principles and phenomenological elements that describe UO control governed by the kidneys, including the glomerular filtration rate (GFR; $J_{GFR}$ in Fig. 1 and Eq. (16)-(18)) modulated by the Starling forces due to the change in the intravascular water and protein volumes (i.e., plasma volume), the reabsorption by the glomerulotubular balance ($J_{RR,GT}$ in Fig. 1 and Eq. (19)-(20)), and the reabsorption and sodium osmosis modulated by the antidiuretic hormone (ADH; $J_{RR,ADH}$ in Fig. 1 and Eq. (19)-(25)). The transient perturbations in volume kinetics and kidney function triggered by burn injury were represented by an array of time-varying phenomenological models acting on various parameters and variables in the mathematical model of volume kinetics to replicate local and systemic pathophysiological changes caused by burn injury, including (i) partial destruction of capillaries in burnt tissues (a in Fig. 1; Eq. (28)-(29)), (ii) denaturation of protein in burnt tissues (b in Fig. 1; Eq. (31)), (iii) transient negative hydrostatic pressure in burnt tissues (c in Fig. 1; Eq. (32)), (iv) increased dermal fluid loss (d in Fig. 1; Eq. (34)-(35)), (v) time-dependent changes in capillary filtration and albumin permeability (e in Fig. 1; Eq. (28)-(30)), and (vi) vasodilation (f in Fig. 1; Eq. 36)). The disruption of the kidney function was the consequence of the perturbations occurring in volume kinetics as governed by Eq. (16)-(26). The equations associated with the mathematical model is summarized in Appendix.

## Clinical Dataset

The clinical dataset used in this paper was furnished from two sources. The first source included 207 burn patients admitted to a burn intensive care unit (ICU) in December 2007-June 2009 [33]. These patients were treated with the aid of a clinical decision support system capable of recommending the hour-by-hour dose of lactated ringer (LR) to maintain UO at a target range of 30-50 ml/hr [10]. The care providers had the authority to override the recommendation. The dataset included hourly UO and LR dose as well as demographics including age, gender, and weight, total burn surface area (TBSA), the presence of inhalation injury, and the time of arrival. The second source included 53 burn patients. 29 patients were treated with the aid of the same clinical decision support system, while 24 patients were treated with the contemporary resuscitation protocols. The dataset included hourly UO and LR dose as well as demographics including age and weight, TBSA, and the time of arrival (gender and presence of inhalation injury were not known). Collectively, age, weight, and TBSA of the patients in the dataset were 47±18 years, 87±22 kg, and 40±18%, respectively. The overall mortality rate of the patients was 30%. In the first source, 77% of the patients were male and 11% of the patients were associated with inhalation injury.

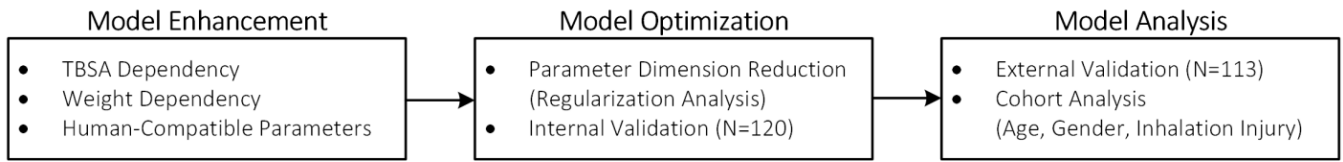

Fig. 2. Analysis procedure. (i) The mathematical model was enhanced to enable in-human application by incorporating TBSA and weight dependence and human-compatible parameter values. (ii) The enhanced mathematical model was optimized to enable its use with limited clinical measurements by model fitting analysis with regularization. The optimized model was internally validated using training dataset consisting of 120 burn patients. (iii) The mathematical model was externally validated using testing dataset consisting of 113 burn patients. The mathematical model was analyzed to garner insights on the pathophysiological differences depending on age, gender, and inhalation injury.

We randomly split the clinical dataset into training dataset to enhance and optimize the mathematical model (N=120) and test dataset (N=113) to validate the optimized mathematical model after excluding 27 burn injury patients associated with prohibitively small number of UO measurements (≤10). The demographic and injury severity of the burn patients in the training and test datasets were comparable (age: 45±19 years versus 49±18 years; weight: 85±18 kg versus 86±22 kg; TBSA 41.5±17.6% versus 38±18). The average hourly UO measurements in the training and test datasets were 23±2 samples and 20±4 samples, respectively.

## Analysis

We conducted the analysis of the clinical dataset to continue to develop, optimize and validate the mathematical model to enable its in-human application (Fig. 2). First, we continued to develop the mathematical model by (i) including TBSA and weight dependency as well as human-compatible parameters to make it globally applicable to burn injury patients associated with diverse demographic characteristics and injury severity ("Model Enhancement" in Fig. 2), and then (ii) optimizing the mathematical model, by systematically reducing the parameter dimension using the training dataset, to make it compatible with sparse clinical measurements ("Model Optimization" in Fig. 2). Second, we validated the optimized mathematical model using the test dataset in terms of its predictive capability and physiological plausibility ("Model Analysis" in Fig. 2). Third, we scrutinized the mathematical models determined specifically for various categorical patient cohorts to gain insights on meaningful pathophysiological characteristics in these categorical patient cohorts ("Model Analysis" in Fig. 2). Details regarding the continued development, optimization, and analysis of the mathematical model are given in II.C.1, II.C.2, and II.C.3 below and are summarized in Table II.

**Mathematical Model Enhancement**

The clinical dataset presents major challenges in estimating all the parameters in the mathematical model on an individual patient basis. First, the burn patients in the clinical dataset exhibit large variability in the demographic characteristics as well as in the severity of burn injury (TBSA ranging between 16% and 100%), both of which increase the inter-individual variability in physiological responses to burn injury and resuscitation. Second, the number of parameters in the mathematical model is excessively large relative to the available measurements (i.e., hourly resuscitation dose and UO are the only measurements available to characterize these burn patients). To address these challenges and seamlessly apply the mathematical model to real burn patients, we leveraged the training dataset to further develop the mathematical model by (i) extending it to accommodate the variability in weight, TBSA, and species as well as (ii) systematically reducing the number of parameters to be estimated using the clinical dataset.

First, we categorized the parameters in the mathematical model into subject-invariant and subject-specific parameters. We categorized as subject-invariant parameters (i) those whose values appear consistent in multiple prior literatures (mostly associated with extensive properties and first principles components in the mathematical model, e.g., nominal water volume and albumin content in the intravascular and tissue compartments and the hydrostatic pressure in Bowman's capsule) and (ii) those whose values must be selected to yield mechanistically relevant physiological responses (e.g., parameters associated with the tissue compliance model, which must be chosen to result in physically relevant tissue hydrostatic pressure for a range of tissue volumes). The values of these subject-invariant parameters were mostly determined based on the existing literature (see Table AI for specific literatures we leveraged to determine these values). We categorized as subject-specific parameters (i) those whose values can exhibit large variability across burn injury patients (e.g., parameters pertaining to burn-induced perturbations and nominal glomerular filtration coefficient), (ii) those whose values have rarely been reported in the literature (e.g., capillary elastance and nominal lymphatic drainage rate), and (iii) those whose values are inherently unknown (e.g., parameters associated with phenomenological elements in the mathematical model). After all, a total of 58 parameters were categorized into 34 subject-invariant ("I" in Table AI) and 24 subject-specific ("S" and "SS" Table AI) parameters.

Second, we improved the mathematical model to accommodate the variability in weight and injury severity in the clinical dataset as well as to increase its suitability to real burn patients. To incorporate the weight dependence into pertinent parameters, we employed a linear allometric relationship by making them linear functions of weight, so that they assume typical values reported in the literature in case of a reference man (70 kg). These parameters include extensive parameters such as the water and protein volumes in the intravascular ($\bar{V}_P$ and $\bar{A}_P$ in Table AI), intact tissue ($\bar{V}_{IT}$ and $\bar{A}_{IT}$ in Table AI), and burnt tissue ($\bar{V}_{BT}$ and $\bar{A}_{BT}$ in Table AI) compartments, capillary filtration rate ($\bar{J}_C$ in Table AI), and lymphatic drainage ($\bar{J}_L$ in Table AI) to list a few. One exception to the linear allometric relationship was the total body surface area ($S_B$ in Table AI), which was made a function of weight through the Haycock formula (Eq. (34c)) and the weight-height relationship reported in the literature [34]. To incorporate the TBSA dependence into pertinent parameters, we (i) made the extensive parameters associated with the burnt tissue compartment functions of TBSA ($\varepsilon_B$ in Table AI) and (ii) expanded the plausible ranges of subject-specific parameter values associated with burn-induced pathophysiological responses so that the estimated parameter values avoid saturation at the pre-specified upper and lower bounds. These parameters include those representing the intensity of the inflammatory responses induced by burn injury such as the maximum increase in the capillary pore radius to albumin radius ratio (i.e. pore ratio) in the intact ($M_{\alpha_{IT}}$ in Table AI) and burnt ($M_{\alpha_{BT}}$ in Table AI) tissue capillary bed, the maximum drop in the burnt tissue hydrostatic pressure ($M_{P_{BT}}$ in Table AI), and the maximum increase in the capillary hydrostatic pressure ($M_{P_C}$ in Table AI) to list a few. To make the mathematical model (which was initially developed based on the dataset collected from animals in our prior work [31]) more compatible with human burn patients, we refined the values of a number of parameters that are inherently different between animals and humans according to the literature (see Table AI for references). These parameters include nominal albumin concentration in the intravascular ([$\bar{A}_P$] in Table AI), intact tissue ([$\bar{A}_{IT}$] in Table AI), and burnt tissue ([$\bar{A}_{BT}$] in Table AI) compartments, colloid oncotic pressure constant ($C_O$ in Table AI), nominal capillary hydrostatic pressure ($\bar{P}_C$ in Table AI), and the total body surface area ($S_B$ in Table AI).

**Mathematical Model Optimization, Training, and Validation**
Using the training dataset, we optimized the mathematical model for in-human use by reducing the number of the subject-specific parameters that must be estimated on an individual patient basis. As described above, we down-selected 24 subject-specific parameters in the mathematical model that must be estimated using the hourly UO and LR dose measurements. Noting that the information content in the hourly UO and LR dose measurements may not be sufficient to robustly estimate all these 24 parameters, we capitalized on the training dataset to split the subject-specific parameters into parameters sensitive versus insensitive to the LR dose-UO input-output relationship. Then, we estimated the sensitive subject-specific parameters on an individual patient basis while fixing the insensitive parameters (together with the 34 subject-invariant parameters) at their typical (i.e., group average) values. First, we determined the typical values of all the 24 subject-specific parameters by fitting the mathematical model to the LR dose-UO measurements pertaining to all the patients in the training dataset based on the pooled approach [35]. This task was accomplished by solving the following optimization problem using a multi-start gradient descent method ("globalsearch" in conjunction with "fmincon") in MATLAB (MathWorks, Natick, MA):

$$\bar{\theta} = arg\,\min_{\theta} \bar{J}(\theta) = arg\,\min_{\theta} \sum_{i=1}^{N} \sqrt{\left(\sum_{k=1}^{D_i} \frac{|uo_i^d(t_k) - uo_i(t_k,\theta)|}{UO_i}\right)^2}, \quad (1)$$

where $\bar{\theta}$ is the vector of typical values of subject-specific parameters, $\theta$ is the vector of 24 subject-specific parameters (i.e., a vector containing the 24 subject-specific parameters in Table AI), $N$ is the number of subjects, $D_i$ is the number of UO measurements for the subject $i$ during the treatment, $uo_i^d(t_k)$ is the value of UO associated with the subject $i$ measured at time $t_k$, $uo_i(t_k, \theta)$ is the value of UO associated with the subject $i$ at time $t_k$ predicted by the mathematical model for a given $\theta$, and $UO_i$ is the normalization factor for UO associated with the subject $i$, which is defined as the range of $uo_i^d$ multiplied by $D_i$. Second, we classified the subject-specific parameters into sensitive and insensitive groups by quantifying and comparing the degree of inter-individual variability associated with all the subject-specific parameters. This task was accomplished by solving the following optimization problem for fitting with regularization [36] on an individual patient basis using a multi-start gradient descent method ("globalsearch" in conjunction with "fmincon") in MATLAB (MathWorks, Natick, MA):

$$\theta_i = arg\,\min_{\theta} J_i(\theta) = arg\,\min_{\theta} \sqrt{\left(\sum_{k=1}^{D_i} \frac{|uo_i^d(t_k) - uo_i(t_k,\theta)|}{UO_i}\right)^2 + \lambda_p \sum_{l=1}^{24} \left|\frac{\theta(l) - \bar{\theta}(l)}{\Theta_l}\right|}, \quad (2)$$

where $\theta_i$ is the vector of 24 subject-specific parameters associated with the subject $i$, $\lambda_p$ is the regularization weight, and $\Theta_l$ is the normalization factor for the $l$-th element $\theta(l)$ of $\theta$, which was defined so that all the elements in $\theta$ are homogeneously ranged approximately between 0 and 1. The regularization discourages the parameters from deviating from their respective typical values unless really needed to achieve superior goodness of fit. Hence, the subset of subject-specific parameters exhibiting deviations from the typical values in many subjects may be viewed as subject-specific parameters sensitive to the LR dose-UO input-output relationship. In this paper, we selected sensitive subject-specific parameters as those whose deviations exceeded a threshold value when averaged across all the 120 patients in the training dataset. Third, we ascertained the ability of the mathematical model (with the chosen sensitive subject-specific parameters) to faithfully replicate the UO responses to the LR dose in the training dataset, as well as its physiological plausibility. To this aim, we estimated the sensitive subject-specific parameters by solving the following optimization problem on an individual patient basis using a multi-start gradient descent method ("globalsearch" in conjunction with "fmincon") in MATLAB (MathWorks, Natick, MA) while fixing the remaining (insensitive subject-specific and subject-invariant) parameters at the respective typical values:

$$\breve{\theta}_i = arg\,\min_{\breve{\theta}} \breve{J}_i(\breve{\theta}) = arg\,\min_{\breve{\theta}} \sqrt{\left(\sum_{k=1}^{D_i} \frac{|uo_i^d(t_k) - y_i(t_k,\breve{\theta})|}{UO_i}\right)^2} \quad (3)$$

where $\breve{\theta}$ is the vector of sensitive subject-specific parameters selected by solving Eq. (2) (i.e., it is a subset of $\theta$), and $\breve{\theta}_i$ is $\breve{\theta}$ estimated for the subject $i$. Then, we examined the faithfulness of the mathematical model in terms of (i) normalized mean absolute error (NMAE) [31], (ii) correlation coefficient, and (iii) UO range agreement, all between measured versus model-replicated UO on an individual patient basis, and (iv) Bland-Altman statistics between all measured versus model-replicated UO. We computed the UO range-based agreement by specifying UO ranges of interest and then for each range computing the percentage of actual UO in the range whose model-predicted UO also resides in the same range. In addition, we examined the physiological plausibility of the mathematical model in terms of (i) both typical and subject-specific model parameter values (e.g., by comparing them with the values reported in the existing literature) as well as (ii) the plausibility of the volume kinetic and kidney function responses predicted by the mathematical model equipped with typical parameter values. We repeated the above procedure to optimize the mathematical model (i.e., Eq. (1)-(3)) so that it can yield minimal number of sensitive subject-specific parameters and adequate faithfulness and physiological plausibility.

Using the test dataset, we externally validated the faithfulness and physiological plausibility of the optimized mathematical model on an individual patient basis, in terms of the same metrics used above.

**Mathematical Model Analysis**

In addition to the optimization and validation of the mathematical model for in-human application (II.C.1-II.C.2), we also sought to garner in-depth insights and expand the knowledge on burn pathophysiology using the mathematical model. In particular, existing literature suggests that patients who are older [37], female [38]–[40], and associated with inhalation injury [37], [41], as well as those who are associated with severe burn injury [37] and/or receive delayed treatment [42] have a higher risk of mortality. The mathematical model already incorporates TBSA and arrival time post-burn, thereby allowing it to predict more severe responses to burn injury associated with large TBSA and delayed resuscitation treatments. However, it does not explicitly account for the effect of age, gender, and inhalation injury.

To investigate if the mathematical model can elucidate the age-, gender-, and inhalation injury-dependent differences in the burn physiological and pathophysiological mechanisms, we fitted the (optimized and validated) mathematical model separately to (i) younger versus older patients, (ii) female versus male patients, and (iii) patients with versus without inhalation injury. We used the patients in the training dataset (N=120), since they were associated with consistent treatment durations (i.e., 24 hours monitoring in most patients) compared to the test dataset. We excluded 16 patients since they did not have gender specification. We defined older patients as those with age above the median age of the 104 patients (45 years), and younger patients otherwise. Using these 104 patients in the training dataset, we built the group-average mathematical models associated with younger (N=52) versus older (N=52) patients, (ii) female (N=22) versus male (N=82) patients, and (iii) patients with (N=11) versus without (N=93) inhalation injury, all by solving a hybrid of the optimization problems in Eq. (1) and Eq. (3) (specifically, solving Eq. (1) only with respect to sensitive subject-specific parameters rather than all the sensitive parameters) based on the dataset associated with the specific patient groups. Then, we examined if the model parameter values for the two groups in each of the three categories (age, gender, and inhalation injury) exhibited meaningful contrasting differences that provide clinically important physiological insights.

# Results

The iterative optimization of the mathematical model using the training dataset resulted in a mathematical model with seven sensitive subject-specific parameters in total, including the nominal capillary pore radius to albumin radius ratio ($\bar{\alpha}$), the maximum increase in the pore ratio in the intact ($M_{\alpha_{IT}}$) and the burnt ($M_{\alpha_{BT}}$) tissue capillary bed, the maximum increase in the capillary hydrostatic pressure ($M_{P_C}$), the slow decay rate associated with the increase in the capillary hydrostatic pressure ($\lambda_{1,P_C}$), the tubule-glomerular feedback sensitivity ($K_{TGF}$), and the nominal water reabsorption rate in the collecting ducts ($\bar{J}_{RR,ADH}$). Table I summarizes NMAE, correlation coefficient, UO range-based agreement pertaining to < 30, 30< <50, and >50 ml/h, or <0.5, 0.5< <1, and >1 ml/kg·hr, and the Bland-Altman statistics (i.e., the limits of agreement), all associated with the optimized mathematical model. Fig. 3 presents examples of actual versus model-predicted UO responses associated with eight patients with various burn injury severity in the test dataset (details discussed in IV.A). Fig. 4 presents volume kinetic and kidney function responses to burn injury and burn resuscitation predicted by the group-average mathematical model in response to group-average burn resuscitation LR dose (details discussed in IV.A). Fig. 5 shows the weight-normalized PV, intravascular water gain (LR dose) and loss (capillary filtration in excess of lymphatic flow), and the burn resuscitation effectiveness (defined as the weight-normalized intravascular water gain rate (i.e., LR dose minus capillary filtration in excess of lymphatic flow) divided by the weight-normalized LR dose) throughout the 24-hour treatment period as predicted by the group-average mathematical model (details discussed in IV.B). Table II summarizes (i) demographics, (ii) statistical characteristics of fluid retention and UO relative to its treatment target range (30-50 ml/hr), and (iii) group-average model parameter values related to burn-induced inflammatory perturbations, all associated with the two patient groups in the three categories.

# Discussion

Developing treatment strategies and expanding knowledge associated with burn injury present formidable challenges due to its complex pathophysiology, large inter-patient variability, and its less common incidence compared to other widespread injuries despite its devastating impact on the mortality and the quality of life. High-fidelity mathematical models capable of replicating volume kinetic and kidney function responses to burn injury and resuscitation has the potential to advance both treatment development and knowledge expansion aspects of burn resuscitation. Regardless, to the best of our knowledge, no mathematical model exists that has been developed and extensively validated using clinical datasets from real burn patients. In this paper, we present our continued development, extensive in-human validation, and analysis of a mathematical model for the study of burn injury and resuscitation, which is equipped with contemporary knowledge on the burn-related physiology and pathophysiology.

### In-Human Credibility

The enhanced/optimized mathematical model based on the training dataset exhibited adequate predictive capability for UO response to burn injury and resuscitation in both training and test datasets (Table I and Fig. 3). In particular, the mathematical model worked equally well in both training and test datasets, both in terms of average statistics and robustness (e.g., NMAE was 15% with small IQR of 6%; Table I). Further, it could capture the physiological differences in burn patients across diverse TBSA range, including those associated with comparable weight ((a) versus (e) and (c) versus (g) in Fig. 3) and those associated with distinct weight ((b) versus (f) and (d) versus (h) in Fig. 3). In addition, the mathematical model showed a high degree of UO range-based agreement (>90% (when weight-normalized) and >78% (when not weight-normalized) of model-predicted UO resided in the same range to actual UO; Table I). Noting that existing burn resuscitation protocols determine the hourly resuscitation dose based on the range of UO, the results suggest that the mathematical model may serve as a valuable platform for non-clinical testing of burn resuscitation protocols and algorithms. In addition to UO, the mathematical model was able to predict the overall volume kinetic and kidney function responses to burn injury and resuscitation in a realistic way: the behaviors of the internal volume kinetic and kidney variables were consistent with the contemporary knowledge on burn pathophysiology as well as findings from recent studies (Fig. 4).

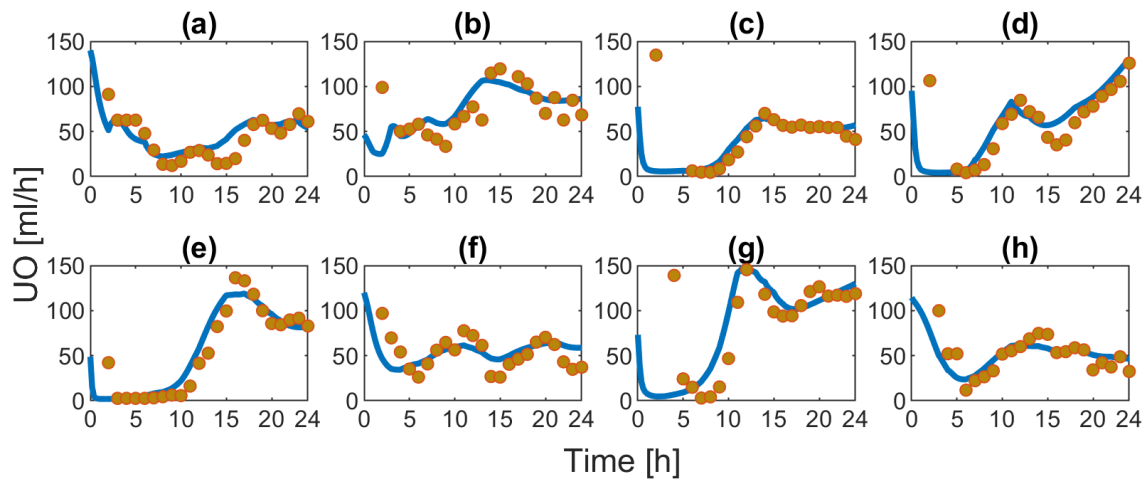

Fig. 3. Actual versus model-predicted urinary output (UO) responses of eight burn patients associated with various injury severity and weight. Circles: actual UO. Solid lines: model-predicted UO. (a) TBSA 27% with 80 kg weight. (b) TBSA 36% with 66 kg weight. (c) TBSA 46% with 90 kg weight. (d) TBSA 60% with 71 kg weight. (e) TBSA 24% with 81 kg weight. (f) TBSA 35% with 94 kg weight. (g) TBSA 50% with 89 kg weight. (h) TBSA 60% with 102 kg weight.

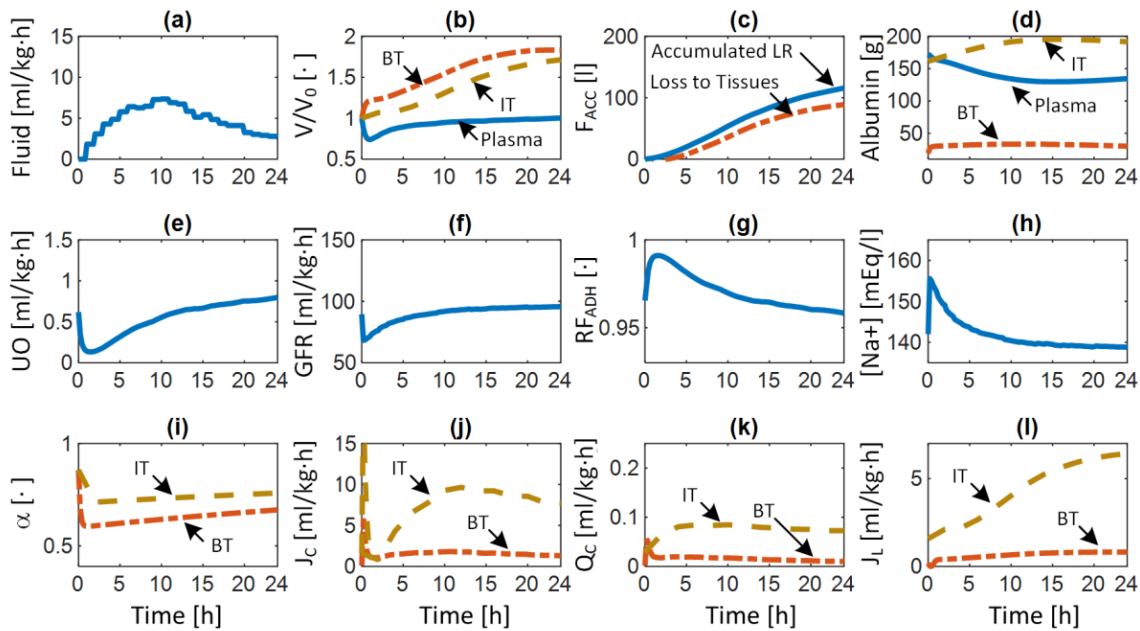

Fig. 4. Volume kinetic and kidney function responses to burn injury and burn resuscitation during initial 24 hours post-burn, predicted by the group-average mathematical model. $V/V_0$: water volume relative to its initial value. $F_{ACC}$: accumulated fluid. $RF_{ADH}$: reabsorption fraction due to ADH (see Appendix B). $\alpha$: Capillary pore radius ratio (see Appendix C). $J_C$: capillary filtration. $Q_C$: albumin transport across the capillary wall. $J_L$: lymphatic flow. (b) and (d): Blue solid, brown dashed, and orange dash-dot lines correspond to plasma, intact tissues, and burnt tissues, respectively. (c): Blue solid and orange dash-dot lines are weight-normalized accumulated resuscitation LR volume and water loss to tissues (i.e., capillary filtration in excess of lymphatic flow), respectively. (i)-(l): Brown dashed, and orange dash-dot lines correspond to intact and burnt tissues, respectively. IT: intact tissues. BT: burnt tissues.

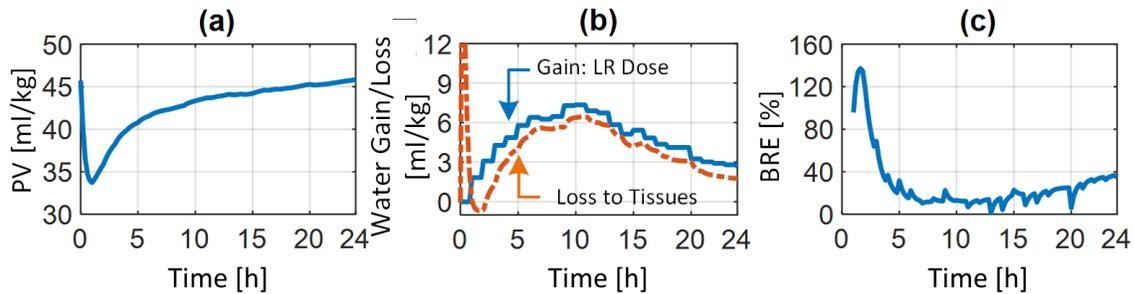

Fig. 5. Group-average prediction of (a) weight-normalized plasma volume (PV), (b) weight-normalized intravascular water gain (LR dose: blue solid) and loss (capillary filtration minus lymphatic flow: orange dashed) rates, and (c) burn resuscitation effectiveness (BRE).

Specifically, the group-average mathematical model predicted that (i) plasma volume and UO showed an anticipated trend of initial decline upon the onset of burn injury and subsequent recovery with the initiation of burn resuscitation and later with the return of resuscitation fluid leaked into the tissues back to the blood (Fig. 4(b), Fig. 4(c), and Fig. 4(e)) [43], [44]; (ii) burnt and intact tissue volumes increased up to nearly twice their initial values, peaking and starting to decay approximately at 24 hours post-burn (Fig. 4(b)) [45]–[47]; (iii) plasma albumin was transported into burnt and intact tissues due to the perturbations in albumin reflection and permeability-surface area coefficients (Fig. 4(d) and Fig. 4(k)) triggered by burn-induced increase in the capillary pore size that decreased the capillary pore radius ratio in both burnt and intact tissue (Fig. 4(i)) [48]; (iv) GFR increased just a few hours post-burn even before plasma volume (Fig. 4(f)) [49]; and (v) sodium concentration decreased after burn injury and resuscitation (Fig. 4(h)) [50].

Importantly, the mathematical model could predict UO as well as physiologically plausible volume kinetic and kidney function responses once physiologically acceptable values were assigned to its parameters. In fact, the majority of the parameters equipped with physiological implications assumed values comparable to typical values and/or those reported in the literature both on the individual and population-average basis (Table AI).

In sum, we demonstrated that the mathematical model can faithfully replicate the volume kinetic and kidney function responses in a wide range of burn patients, both in terms of the adequacy of the model-predicted responses and the plausibility of the model parameter values.

## Insights on Burn Resuscitation Effectiveness

One strength of the mathematical model presented in this paper is its ability to replicate overall responses of a burn patient to injury and resuscitation, including those that cannot be clinically measured. Exploiting this advantage, we sought to garner insights on the effectiveness of burn resuscitation in a typical patient subject to burn injury. It is known that the homeostasis in volume kinetics is severely disrupted after burn injury due to the activation of multiple inflammatory mediators, which in turn causes a large portion of the resuscitation fluid to leak out of the intravascular compartment via capillary filtration. Although this leakage is partially recovered by the increase in lymphatic flow, >50% of the resuscitation fluid can leak out of the intravascular compartment in the initial hours post-burn in extensive burn injury [51], [52]. In this regard, burn resuscitation effectiveness represents the portion of the resuscitation fluid actually used to expand plasma volume. Based on the investigation and interpretation of >50 physiological variables including those presented in Fig. 4 and Fig. 5, we could garner the following insights on the important physiological and pathophysiological mechanisms responsible for the effectiveness of burn resuscitation during the initial 24 hours post-burn. Initially, there is a large and fast fluid shift from the blood to the intact and burnt tissues immediately after burn injury for up to one hour, leading to a large decrease in the plasma volume (Fig. 4(j) and Fig. 5(a)). Our mathematical model suggests that major mechanisms responsible for this initial loss of plasma volume may be negative hydrostatic pressure, protein denaturation in burnt tissues, and systemic increase in the capillary hydrostatic pressure.

After this initial phase, a decrease in the capillary filtration and the self-regulation of plasma volume occurs for up to one hour. Our mathematical model suggests that major mechanisms responsible for this phase may be the reduction in the plasma volume and the resulting decrease in the capillary hydrostatic pressure, the recovery of hydrostatic pressure in the burnt tissues, and the increase in the lymphatic flow (Fig. 4(l)). The effectiveness of burn resuscitation during this phase is very high (>100%), meaning that plasma volume is expanded based on almost all the resuscitation fluid as well as the fluid returning from the edematous (burnt and intact) tissues (Fig. 5(c)). Subsequently, burn resuscitation effectiveness is deteriorated quickly as the capillary filtration of water and protein increases again due to the opened capillary pores (Fig. 4(i)) and the increase in the plasma volume. Our mathematical model suggests that major mechanisms responsible for this phase may include the increase in the protein concentration in both burnt and intact tissues as well as the hypoproteinemia in the blood (Fig. 4(d)), which altogether increase the osmotic pressure gradient toward tissues and promote capillary filtration of both water and protein (thereby forming a vicious circle). Burn resuscitation effectiveness reaches its minimum level of 2%-15% at 10-15 hours after the initiation of treatment, which is in close agreement with the literature suggesting maximal edema formation in this period post-burn [45]. Finally, burn resuscitation effectiveness increases back to approximately 40% at 24 hours after the initiation of treatment. Our mathematical model suggests that major mechanisms responsible for this recovery may include the recovery of lymphatic flow to return excessive water and protein to the blood as well as the gradual decrease in the capillary pore size, which altogether decreases the fluid extravasation rate.

## Cohort-Dependent Differences in Burn Pathophysiology

The analysis of datasets associated with various categorical patient cohorts (with respect to age, gender, and inhalation injury) provided meaningful insights on the cohort-dependent differences in burn physiology and pathophysiology (Table II). To begin with, the mathematical model was able to replicate UO response to burn injury and resuscitation associated with all the categorical patient cohorts (younger versus older, female versus male, and patients with and without inhalation injury).

First, between younger versus older patients, the latter had much higher mortality rate and higher portion of UO responses below the target therapeutic range than the former despite its smaller group-average TBSA and higher level of weight-normalized LR dose, which leads to higher fluid retention in the latter (Table II, first two columns). Comparing the mathematical models fitted to younger versus older burn patients, the latter was associated with higher inflammation factors (including the larger increase in capillary pore size in the burnt tissues ($M_{\alpha_{BT}}$) and the capillary hydrostatic pressure ($M_{P_C}$)). Second, between patients with versus without inhalation injury, the former likewise had >1.3 times higher mortality rate and higher portion of UO responses below the target therapeutic range than the latter (Table II, last two columns). Comparing the mathematical models fitted to burn patients with versus without inhalation injury, the former exhibited higher inflammation factors (including the larger increase in capillary pore size ($M_{\alpha_{IT}}$ and $M_{\alpha_{BT}}$) and capillary hydrostatic pressure ($M_{P_C}$)) similarly to the older patient cohort. Existing literature shows the possible association between fluid retention and edema versus mortality and complication rates in burn patients [12]. In addition, both literature and our dataset indicate higher fluid retention and edema as well as higher mortality rates in older patients and patients with inhalation injury [53], [54]. From this standpoint, our mathematical model analysis predict that higher inflammation may be a key contributing factor in increasing the mortality risk of older patients and patients with inhalation injury via an increase in the fluid retention.

Table II: Demographics, characteristics of fluid resuscitation, fluid retention, and urinary output (UO) relative to its treatment target range (30-50 ml/hr), and group-average model parameter values related to burn-induced inflammatory perturbations, all associated with the two patient groups in age (younger versus older patients), gender (female versus male patients), and inhalation injury (patients with versus without inhalation injury) categories. Fluid retention is computed as the total resuscitation fluid (LR) volume minus the total UO during the 24 hours of treatment. Fluid resuscitation and fluid retention are shown in mean±SD, while UO is shown in median (IQR). $M_{\alpha_{BT}}$, $M_{\alpha_{IT}}$, and $M_{P_C}$ are group-average values of burn-induced inflammatory perturbation parameters in the mathematical model.

| | Younger (N=52) | Older (N=52) | Female (N=22) | Male (N=82) | Injury (N=11) | No Injury (N=93) |
|---|---|---|---|---|---|---|
| Weight [kg] | 86 (21) | 84 (15) | 74 (17) | 88 (17) | 77 (15) | 86 (18) |
| Injury Severity (TBSA) [%] | 44 (17) | 38 (17) | 38 (13) | 43 (19) | 46 (14) | 41 (18) |
| Mortality Rate [%] | 18 | 42 | 37 | 27 | 40 | 29 |
| Fluid Resuscitation [ml/kg·%] | 3.27±1.18 | 4.20±1.59 | 4.0±1.24 | 3.67±1.52 | 3.66±1.49 | 4.33±1.19 |
| Fluid Retention [ml/kg·%] | 2.78±1.16 | 3.75±1.6 | 3.43±1.2 | 3.2±1.5 | 3.0±1.5 | 3.9±1.2 |
| 30 ml/hr≤UO≤50 ml/hr | 26 (15) | 23 (16) | 28 (14) | 24 (16) | 24 (15) | 24 (16) |
| UO<30 ml/hr | 21 (14) | 34 (20) | 32 (15) | 26 (20) | 27 (18) | 31 (23) |
| UO>50 ml/hr | 53 (18) | 42 (20) | 40 (16) | 50 (20) | 48 (20) | 44 (19) |
| $M_{\alpha_{BT}}$ | 0.23 | 0.54 | 0.32 | 0.48 | 0.31 | 0.52 |
| $M_{\alpha_{IT}}$ | 0.19 | 0.19 | 0.16 | 0.15 | 0.18 | 0.19 |
| $M_{P_C}$ | 0.05 | 1.70 | 2.26 | 1.73 | 0.13 | 1.47 |

Our prediction is in fact consistent with the contemporary knowledge in the literature identifying inflammation as an important mediator of increased fluid retention and edema with the increased mortality rate in older patients [55], [56] and patients with inhalation injury [57], [58], although other causes can play a role (e.g., degraded cardiovascular efficiency in elderly burn patients [59]).

Although our mathematical model analysis reveals possible mechanisms responsible for higher mortality rate, the exact cause is yet to be clearly elucidated. Regardless, lower burn resuscitation effectiveness in older patients and patients with inhalation injury relative to younger patients and patients without inhalation injury remains true, and our mathematical model was able to replicate the age- and inhalation injury-dependent differences in burn resuscitation effectiveness. Hence, our mathematical model may serve as an effective basis to develop and validate burn resuscitation protocols and algorithms suited to these categorical patient cohorts.

Third, between female and male patients, neither the dataset nor the mathematical model showed any meaningful difference in terms of inflammation and fluid retention (Table II, two middle columns). This contrasts against some literature identifying the female gender as a mediator of mortality risk associated with burn injury, as confirmed by our dataset (1.3 times the mortality rate in males). Hence, our mathematical model analysis suggests that higher mortality risk in female burn patients may be attributed to factors other than an increase in the inflammation and the corresponding increase in the fluid retention, especially those not manifested in the initial 24 hours post-burn. In fact, a prior study performed on mice showed that the difference in the inflammatory responses in female and male subjects was not clear until 6 days post-burn [60]. The exact mechanisms responsible for the gender difference in burn-induced mortality risk are still unknown and must be unveiled.

# Conclusions

The main outcome of this paper is a physiologically plausible mathematical model capable of replicating volume kinetic and kidney function responses to burn injury and resuscitation suited to in-human application. To the best of our knowledge, the mathematical model presented in this paper may be the first of its kind developed and extensively validated using large clinical datasets from real burn patients. We anticipate that the mathematical model may provide an attractive platform to conduct non-clinical testing of burn resuscitation protocols and test new hypotheses on burn pathophysiology. Future effort must be exerted to investigate the potential of the mathematical model as medical digital twin for disciplined development and rigorous stress testing of emerging burn resuscitation algorithms and as a cornerstone to expand our understanding of burns.

# Appendix

This section summarizes the mathematical model equations and parameter values. Full complete details on the mathematical model can be found in our prior work [31] and the references therein.

## Volume Kinetics

Volume kinetics is represented by a classical multi-compartmental model consisting of vasculature, intact tissues, and burnt tissues as separate compartments. The water (Eq. (4)) and albumin (Eq. (5)) balances in each compartment are formulated based on the mass conservation principle:

$$\frac{d(V_P)}{dt} = -J_{C,BT} - J_{C,IT} + J_{L,BT} + J_{L,IT} + J_F - J_U, \tag{4a}$$

$$\frac{d(V_{BT})}{dt} = J_{C,BT} - J_{L,BT} - J_{EX} - J_{EV,BT}, \tag{4b}$$

$$\frac{d(V_{IT})}{dt} = J_{C,IT} - J_{L,IT} - J_{EV,IT}, \tag{4c}$$

$$\frac{d(A_P)}{dt} = -Q_{C,BT} - Q_{C,IT} + Q_{L,BT} + Q_{L,IT} + Q_F, \tag{5a}$$

$$\frac{d(A_{BT})}{dt} = Q_{C,BT} - Q_{L,BT} - Q_{EX} + Q_{PD}, \tag{5b}$$

$$\frac{d(A_{IT})}{dt} = Q_{C,IT} - Q_{L,IT}, \tag{5c}$$

where $V$ is water volume, $A$ is albumin content, $J$ is water flow, and $Q$ is albumin flow. The subscripts $C$, $L$, $F$, $U$, $EV$, $EX$, and $PD$ denote capillary filtration, lymphatic flow, fluid infusion, UO, evaporation, exudation, and protein influx due to burn-induced denaturation (see Appendix C), respectively, while the subscripts P, BT, IT denote plasma, burnt tissues, and intact tissues. The capillary filtration is expressed using the Starling equation:

$$J_{C,X} = K_{C,X}[P_C - P_X - \sigma_X(\pi_C - \pi_X)], \tag{6}$$

$$\pi_C = C_O[A]_P, \; \pi_X = C_O[A]_X, \; [A]_P = \frac{A_P}{V_P}, \; [A]_X = \frac{A_X}{V_X}, \tag{7}$$

where $X \in \{BT, IT\}$, $P_C$ and $P_X$ are capillary and tissue hydrostatic pressures, $\pi_C$ and $\pi_X$ are plasma and tissue colloid oncotic pressures, and $K_{C,X}$ and $\sigma_X$ are capillary filtration and albumin reflection coefficients, respectively. $[A]_P$ and $[A]_X$ are plasma and tissue albumin concentrations, respectively. $C_O$ is the colloid oncotic pressure constant. The lymphatic water flow is expressed by a phenomenological model in the form of a sigmoidal curve [31]:

$$J_{L,X} = \frac{\bar{J}_{L,X}}{C_L + (1 - C_L)e^{-S_L(P_X - \bar{P}_X)}}, \tag{8}$$

where $\bar{J}_{L,X}$ is nominal lymphatic flow at nominal tissue hydrostatic pressure $\bar{P}_X$, and $C_L$ and $S_L$ are constants representing the inverse of the maximal degree of increase in the lymphatic flow and its sensitivity to the change in the tissue hydrostatic pressure. The lymphatic albumin flow depends both on the lymphatic water flow and the albumin concentration of the corresponding tissue:

$$Q_{L,X} = J_{L,X}[A]_X. \tag{9}$$

The albumin transport across the capillary wall is expressed based on the coupled diffusion-convection equation [61], where PS is permeability-surface area coefficient:

$$Q_{C,X} = J_{C,X}(1 - \sigma_X) \left\{ \frac{[A]_P - [A]_X e^{-\frac{(1-\sigma_X)J_{C,X}}{PS_X}}}{1 - e^{-\frac{(1-\sigma_X)J_{C,X}}{PS_X}}} \right\}. \tag{10}$$

A linear phenomenological relationship between $P_C$ and $V_P$ is assumed:

$$P_C = \bar{P}_C + E_C(V_P - \bar{V}_P), \tag{11}$$

where $\bar{P}_C$ is nominal capillary hydrostatic pressure associated with the nominal PV ($\bar{V}_P$), and $E_C$ is capillary elastance. For the interstitial compartments, we use the nonlinear microscopic pressure-volume model developed by Øien and Wiig [62] and employ it in our macroscopic context:

$$P_X = -\frac{\alpha}{R} + \gamma, \tag{12}$$

$$R = R(y_X) = \hat{R}\left[1 - (1 - \beta)\left(\frac{y_X - \hat{y}}{\check{y} - \hat{y}}\right)\right]^n, \tag{13}$$

$$y_X = 0.75\left(1 + \bar{W}_X \frac{V_X}{\bar{V}_X}\right), \tag{14}$$

where $\alpha$, $\gamma$, $\hat{R}$, $\beta$, and n are constant coefficients (Table AI), $y_X$ is a measure of hydration in the tissue $X$ at the microscopic level, and $\hat{y}$ and $\check{y}$ its maximum and minimum values, and $R$ is the radius of glycosaminoglycans, varying in response to $y_X$. Eq. (11) relates the microscopic hydration to macroscopic hydration, where $\bar{W}_X$ is nominal hydration level, defined as the ratio of the water volume in $X$ and its dry weight [30].

## Kidney Function

The renal function is based on by a novel lumped-parameter, hybrid mechanistic-phenomenological model we developed in our previous work [31]. As blood circulates in the body, the renal plasma flow ($J_{RPF}$) flows into the kidneys to be filtered. The glomerular filtration rate ($J_{GFR}$) is the fraction of $J_{RPF}$ that is filtered into the Bowman's capsule to flow along the nephron tubules, the majority of which is reabsorbed back into the circulation, cleared from waste material. UO is given by the difference between glomerular filtration rate ($J_{GFR}$) and reabsorption rate (RR; $J_{RR}$):

$$J_U = J_{GFR} - J_{RR}. \tag{15}$$

$J_{GFR}$ is derived from the Starling forces between glomerular capillaries and Bowman's capsule as follows [31]:

$$J_{GFR} = K_G[P_G - P_B - \pi_G] \tag{16.a}$$

$$P_G = \lambda_G \frac{J_{RPF}}{R_G}, \tag{16.b}$$

where $\lambda_G$ is a constant coefficient relating the glomerular hydrostatic pressure ($P_G$) to the glomerular resistance ($R_G$) and $J_{GFR}$, and $K_G$ is the glomerular filtration coefficient. The hydrostatic pressure in Bowman's capsule ($P_B$) is assumed constant during burn injury and resuscitation [63]. Assuming that the albumin flowing into the kidney does not pass into the Bowman's capsule, we can write:

$$[A]_C J_{RPF} = [A]_G (J_{RPF} - J_{GFR}) \rightarrow [A]_G = \frac{J_{RPF}}{J_{RPF} - J_{GFR}}[A]_C = \frac{1}{1 - \varepsilon_{GFR}}[A]_C, \tag{16.c}$$

where $[A]_G$ is the colloid content of the glomeruli, and $\varepsilon_{GFR} = \frac{J_{GFR}}{J_{RPF}}$ is the filtration fraction. Since $\varepsilon_{GFR}$ is naturally always less than 1, assuming the colloid oncotic pressure constant in the glomeruli is identical to other capillaries:

$$\pi_G = \frac{1}{1 - \varepsilon_{GFR}} \pi_C \approx (1 + \varepsilon_{GFR})\pi_C, \tag{16.d}$$

Combining Eq. (16.1)-(16.4) we get:

$$J_{GFR} \approx \frac{K_G J_{RPF}}{J_{RPF} + K_G \pi_C}\left[\lambda_G \frac{J_{RPF}}{R_G} - P_B - \pi_C\right], \tag{16.e}$$

$J_{RPF}$ is perturbed by the fluctuations in cardiac output, which in our mathematical model is assumed to be proportional to PV:

$$J_{RPF} = \bar{J}_{RPF} \frac{V_P}{\bar{V}_P}, \tag{17}$$

where $\bar{J}_{RPF}$ is nominal $J_{RPF}$ corresponding to nominal plasma volume $\bar{V}_P$. The perturbation in $J_{RPF}$ is strictly compensated for by the modulation of renal capillary resistance via tubulo-glomerular feedback (TGF). We express TGF as simple phenomenological dynamic modulation of $R_G$ which is a hypothetical variable representing glomerular resistance [31]:

$$\tau_{TGF}\frac{d\Delta R_G}{dt} = -\Delta R_G + \frac{K_{TGF}}{\bar{J}_{GFR}}(J_{GFR} - \bar{J}_{GFR}), \tag{18}$$

where $\Delta R_G = R_G - \bar{R}_G$ is the deviation of $R_G$ from its nominal value $\bar{R}_G$ enforced by TGF, $\tau_{TGF}$ and $K_{TGF}$ are time constant and sensitivity associated with TGF, and $\bar{J}_{GFR}$ is the nominal value of $J_{GFR}$. Combining Eq. (16)-(18), $J_{GFR}$ is expressed in terms of PV and plasma albumin content.

$J_{RR}$ is assumed to be modulated by the glomerulotubular balance (GT, $J_{RR,GT}$), and pure water reabsorption in the collecting ducts governed by the Antidiuretic hormone levels (ADH, $J_{RR,ADH}$):

$$J_{RR} = J_{RR,GT} + J_{RR,ADH}. \tag{19}$$

GT modulates proximal tubules so that approximately 80% of GFR is reabsorbed [64], [65]:

$$J_{RR,GT} \approx 0.8 J_{GFR}. \tag{20}$$

ADH content is modulated by baroreceptor (inversely related to the changes in PV) and osmoreceptor (directly related to the changes in sodium concentration ([$Na^+$]) in the blood) signals [66], [67]. Typically, 27% of ADH is excreted without reabsorption via GFR, while 73% is metabolized in the liver proportionally to blood flow (which is assumed to vary with PV). The dynamics of the ADH content may then be expressed as follows [31]:

$$\frac{d(ADH)}{dt} = K_{ADH} e^{\left(-\lambda_{V_P} \Delta V_P + \lambda_{[Na^+]} \Delta [Na^+]\right)} - 0.27 K_{ADH} [ADH] \frac{J_{GFR}}{\bar{J}_{GFR}} - 0.73 K_{ADH} [ADH] \frac{V_P}{\bar{V}_P}, \tag{21}$$

where $ADH$ is ADH content in the extracellular fluid, $K_{ADH}$ is nominal ADH secretion rate, $\lambda_{V_P}$ and $\lambda_{[Na^+]}$ are positive constant coefficients, and $[ADH]$ is ADH concentration in the extracellular fluid:

$$[ADH] = \frac{ADH}{V_P + V_{BT} + V_{IT}}. \tag{22}$$

The nominal ADH value is chosen such that the corresponding nominal ADH concentration ($\overline{[ADH]}$) equals 1 [pg/ml], because the absolute value of ADH concentration is not important in our context. ADH modulates the reabsorption fraction of the fluid reaching the collecting ducts ($\varepsilon_{RR}$). We adopt the Michaelis-Menten equation to express the relationship between $[ADH]$ and the $\varepsilon_{RR}$ [67]:

$$\varepsilon_{RR} = K_{RR} \frac{[ADH]}{[ADH]_{50} + [ADH]}, \tag{23}$$

$$J_{RR,ADH} = \varepsilon_{RR}(J_{GFR} - J_{RR,GT}), \tag{24}$$

where $K_{RR}$ is maximum $\varepsilon_{RR}$, and $[ADH]_{50}$ is ADH concentration corresponding to $\varepsilon_{RR} = \frac{1}{2} K_{RR}$. Based on the idealistic assumption that sodium content is conserved in the body (including the collecting ducts), and is diluted as the pure water reabsorption increases in the collecting ducts [67]:

$$[Na^+] = \frac{\bar{J}_{RR,ADH}}{J_{RR,ADH}} \overline{[Na^+]}, \tag{25}$$

where $\overline{[Na^+]}$ is nominal sodium concentration corresponding to $\bar{J}_{RR,ADH}$. Combining Eq. (15)-(25), $J_U$ is expressed in terms of plasma volume, plasma albumin content, and ADH concentration:

$$J_U = J_{GFR} - J_{RR} = J_{GFR} - J_{RR,GT} - J_{RR,ADH} \approx 0.2 \frac{K_G J_{RPF}(V_P)}{J_{RPF} + K_G \pi_C} \left[ \lambda_G \frac{J_{RPF}(V_P)}{R_G} - P_B - \pi_C \right] (1 - \varepsilon_{RR}). \tag{26}$$

## 1.1 BURN-INDUCED PATHOPHYSIOLOGY IN VOLUME KINETICS AND KIDNEY FUNCTION

To express various perturbations induced by burn injury, a universal perturbation function below is used:

$$\phi(M_W, \lambda_{1,W}, \lambda_{2,W}, t) = M_W \left( e^{-\lambda_{1,W} t} - e^{-\lambda_{2,W} t} \right), \tag{27}$$

where $M_W$ is maximum perturbation, and $\lambda_{1,W}$ and $\lambda_{2,W}$ are slow and fast time constants, all corresponding to a perturbation $W \in \{\alpha_{BT}, \alpha_{IT}, P_C, P_{BT}\}$, where $\alpha_{BT}$ and $\alpha_{IT}$ are pore radius ratios associated with burnt and intact tissues (see below for details). Eq. (27) describes an intensifying perturbation post-burn, which reaches its maximum at a varying time post-injury and then starts to decay. It can be incorporated into volume kinetics and kidney function models above to replicate burn-induced perturbations. Details follow.

The destruction of capillaries in burnt tissues is expressed as perturbations in capillary filtration coefficient $K_{C,BT}$ and permeability surface area coefficient $PS_{BT}$. Based on the pore theory of trans-capillary exchange [68], [69], capillary filtration, albumin reflection, and permeability-surface area coefficients associated with burnt and intact tissues are expressed as functions of the pore ratios $\alpha_X$, $X \in \{BT, IT\}$, defined as the ratio between albumin radius and capillary pore radius:

$$K_{C,X} = \overline{K}_{C,X} k_{PD,X} \frac{\bar{\alpha}_X^4}{\alpha_X^4}, \tag{28a}$$

$$\sigma_X = 1 - (1 - \alpha_X)^2, \tag{28b}$$

$$PS_X = \overline{PS}_X k_{PD,X} \frac{\bar{\alpha}_X^2(1-\alpha_X^2)}{(1-\bar{\alpha}_X^2)\alpha_X^2}, \tag{28c}$$

where $\overline{K}_{C,X}$, $\overline{PS}_X$, and $\bar{\alpha}_X$ are nominal values of $K_{C,X}$, $PS_X$, and $\alpha_X$ adjusted for the water fraction in burnt and intact tissues, weight, and capillary recruitment [21]:

$$\overline{K}_{C,BT} = \overline{K}_C \varepsilon_B r_{FV} \eta_{CR}, \quad \overline{K}_{C,IT} = \overline{K}_C (1 - \varepsilon_B r_{FV}) \eta_{CR}, \tag{29a}$$

$$\overline{PS}_{BT} = \overline{PS} \varepsilon_B r_{FV} \eta_{CR}, \quad \overline{PS}_{IT} = \overline{PS}(1 - \varepsilon_B r_{FV}) \eta_{CR}, \tag{29b}$$

where $\overline{K}_C$ and $\overline{PS}$ are nominal capillary filtration and permeability surface area coefficients in the absence of burn injury, $\eta_{CR} = \left(2\frac{V_P}{\bar{V}_P} - 1\right)$ (which shows that the parameters in Eq. (29) are extensive and depend on PV), and $r_{FV}$ is the fluid volume ratio between skin and total interstitial compartment [21]. Burn-induced changes in these coefficients are expressed by formalizing the burn-induced changes in the capillary pore radius ratios using Eq. (27):

$$\alpha_X(t) = \bar{\alpha}_X - \phi(M_{\alpha_X}, \lambda_{1,\alpha_X}, \lambda_{2,\alpha_X}, t). \tag{30}$$

The denaturation of protein in burnt tissues ($Q_{PD}$ in Eq. (5b)) is expressed as a protein influx $Q_{PD}$ into burnt tissues [28], [48], [70]:

$$Q_{PD} = \hat{Q}_{PD} e^{-\lambda_{PD} t}, \tag{31}$$

where $\hat{Q}_{PD}$ is protein influx immediately post-burn, decaying with a time constant $\lambda_{PD}$.

The transient negative hydrostatic pressure in burnt tissues is expressed using Eq. (27):

$$\Delta P_{BT}(t) = -\phi(M_{P_{BT}}, \lambda_{1,P_{BT}}, \lambda_{2,P_{BT}}, t). \tag{32}$$

The overall hydrostatic pressure in burnt tissues is computed by combining Eq. (12) and Eq. (32):

$$P_{BT}(t) = -\frac{\alpha}{R(y_{BT})} + \gamma + \Delta P_{BT} = -\frac{\alpha}{R(y_{BT})} + \gamma - \phi(M_{P_{BT}}, \lambda_{1,P_{BT}}, \lambda_{2,P_{BT}}, t). \tag{33}$$

Evaporation and exudation (Eq. 4(b)-4(c)) are modeled based on the literature [71], [72]:

$$J_{EV,BT} = \begin{cases} K_{1,EV} \varepsilon_B S_B e^{\lambda_{1,EV} t}, & t < 6 \, hr \\ K_{2,EV} \varepsilon_B S_B e^{\lambda_{2,EV} t}, & t > 6 \, hr \end{cases}, \tag{34a}$$

$$J_{EV,IT} = K_{1,EV}(1-\varepsilon_B)S_B, \tag{34b}$$

$$S_B = 0.0242 H^{0.396} W^{0.538}, \tag{34c}$$

$$J_{EX} = K_{EX}\varepsilon_B S_B e^{\lambda_{EX}t}, \tag{35a}$$

$$Q_{EX} = J_{EX}\eta_{EX}[A]_{BT}, \tag{35b}$$

where $K_{1,EV}, K_{2,EV}, K_{EX}, \lambda_{1,EV}, \lambda_{2,EV}$, and $\lambda_{EX}$ are constant coefficients (Table AI), $\varepsilon_B$ is the fraction of body surface area subject to burn, $S_B$ is total body surface area based on the Haycock formula, and $\eta_{EX}$ is the ratio between albumin concentration in the exudate and albumin concentration in the burnt tissues [21].

The perturbation in capillary hydrostatic pressure is expressed using Eq. (27) [73]:

$$\Delta P_C(t) = \phi(M_{P_C}, \lambda_{1,P_C}, \lambda_{2,P_C}, t). \tag{36}$$

The overall capillary hydrostatic pressure is computed by combining Eq. (11) and Eq. (36):

$$P_C = \bar{P}_C + E_C(V_P - \bar{V}_P) + \Delta P_C(t) = \bar{P}_C + E_C(V_P - \bar{V}_P) + \phi(M_{P_C}, \lambda_{1,P_C}, \lambda_{2,P_C}, t). \tag{37}$$

## Model Parameters: Nomenclature, Definitions, and Values

Table AI: Mathematical model parameters: definitions, categories (I/S/SS), and values. I: subject-invariant parameters. S: subject-specific parameters. SS: sensitive subject-specific parameters. Parameter values are given as mean, median (IQR), or mean+/-SD.

| Symbol | Definition | I/S | Value (Model) | Value (Literature) |
|---|---|---|---|---|
| $\bar{V}_P$ | Nominal water volume in plasma [ml/kg] | I | 42.8 | 42 [74]-46 [21] |
| $r_{FV}$ | Skin fluid volume to total interstitial fluid volume ratio [·] | I | 0.28 | 0.28 [21] |
| $\bar{V}_{BT}$ | Nominal water volume in burnt tissue [ml/kg] | I | $120\varepsilon_B r_{FV}$ | $120\varepsilon_B r_{FV}$ [21] |
| $\bar{V}_{IT}$ | Nominal water volume in intact tissue [ml/kg] | I | $120(1-\varepsilon_B r_{FV})$ | $120(1-\varepsilon_B r_{FV})$ [21] |
| $[\bar{A}_P]$ | Nominal albumin concentration in plasma [g/ml] | I | 0.045 | 0.035-0.045 [75] |
| $[\bar{A}_{BT}]$ | Nominal albumin concentration in burnt tissue [g/ml] | I | 0.018 | 0.013 [76]-0.016 [21] |
| $[\bar{A}_{IT}]$ | Nominal albumin concentration in intact tissue [g/ml] | I | 0.018 | 0.013 [76] -0.016 [21] |
| $\bar{A}_P$ | Nominal albumin content in plasma [g] | I | $[\bar{A}_P]\bar{V}_P$ | - |
| $\bar{A}_{BT}$ | Nominal albumin content in burnt tissue [g] | I | $[\bar{A}_{BT}]\bar{V}_{BT}$ | - |
| $\bar{A}_{IT}$ | Nominal albumin content in intact tissue [g] | I | $[\bar{A}_{IT}]\bar{V}_{IT}$ | - |
| $\bar{J}_C$ | Nominal capillary filtration [ml/kg·h] | S | 1.76 | 1.72 [20] |
| $C_O$ | Colloid oncotic pressure constant [mmHg/g·ml] | I | 609 | 657 [21] |
| $\bar{J}_L$ | Nominal total lymphatic flow to plasma [ml/kg·h] | I | 1.76 | 1.07 [32]-2.9 [77] |
| $\bar{J}_{L,BT}$ | Nominal lymphatic flow from burnt tissue to plasma [ml/kg·h] | I | $\bar{J}_L\varepsilon_B r_{FV}$ | $\bar{J}_L\varepsilon_B r_{FV}$ [21] |
| $\bar{J}_{L,IT}$ | Nominal lymphatic flow from intact tissue to plasma [ml/kg·h] | I | $\bar{J}_L(1-\varepsilon_B r_{FV})$ | $\bar{J}_L(1-\varepsilon_B r_{FV})$ [21] |
| $C_L$ | Lymphatic maximal increase coefficient [·] | S | 0.07 | - |
| $S_L$ | Lymphatic pressure sensitivity coefficient [1/mmHg] | S | 1.2 | - |
| $\bar{P}_C$ | Nominal capillary hydrostatic pressure [mmHg] | S | 16.3 | 13 [78]- 24 [45] |
| $E_C$ | Capillary elastance [mmHg/ml] | S | 0.0084 | 0.0097 [32] |
| $\alpha$ | Tissue electrostatic pressure coefficient [mmHg] | I | 10 | 10 [62] |
| $\gamma$ | Tissue tension pressure coefficient [mmHg] | I | 3.75 | 3.75 [62] |
| $\hat{y}$ | Maximum half-thickness of the extracellular matrix [·] | I | 4 | 4 [62] |
| $\check{y}$ | Minimum half-thickness of the extracellular matrix [·] | I | 1 | 1 [62] |
| $\hat{R}$ | Maximum GAG radius [·] | I | 3.5 | 3.5 [62] |
| $\beta$ | Radius threshold ratio [·] | I | 0.23 | 0.23 [62] |
| $n$ | Hydration response coefficient [·] | I | 8 | 2-8 [62] |
| $\bar{W}_X$ | Nominal hydration level [ml/g] | I | 0.66 | 0.23-0.81 [30], [62] |
| $\bar{J}_{RPF}$ | Nominal renal plasma flow [ml/kg·h] | I | 536 | 536 [79] |
| $\tau_{TGF}$ | Tubuglomerular feedback time constant [1/h] | S | 0.35 | - |

| Symbol | Description | Type | Value | Reference |
|---|---|---|---|---|
| $K_{TGF}$ | Tubuglomerular feedback sensitivity [·] | SS | 1.45 (0.4) | - |
| $\bar{R}_G$ | Nominal glomerular resistance [mmHg/ml/kg·h] | I | 9.05 | - |
| $\lambda_G$ | Glomerular hydrostatic pressure sensitivity [mmHg$^2$/(ml/h·kg)$^2$] | I | 1 | - |
| $P_B$ | Hydrostatic pressure in Bowman's capsules [mmHg] | I | 18 | 18 [63] |
| $K_G$ | Glomerular filtration coefficient [ml/kg·h·mmHg] | S | 9.2 (1.2) | 9.6-12.0 [80] |
| $K_{ADH}$ | Nominal ADH secretion rate [pg/kg·h] | I | 287 | 287 [81], [82] |
| $\lambda_{V_P}$ | ADH sensitivity to plasma volume change [1/ml] | S | 0.0011 | - |
| $\lambda_{[Na^+]}$ | ADH sensitivity to sodium concentration change [l/mEq] | S | 0.085 | - |
| $K_{RR}$ | Maximum collecting duct reabsorption fraction [·] | I | 0.999 | - |
| $\overline{[ADH]}$ | Nominal ADH concentration [pg/ml] | I | 1 | 0-5 [83] |
| $[ADH]_{50}$ | ADH concentration corresponding to $\frac{1}{2}K_{RR}$ [pg/ml] | S | 0.0594 | - |
| $\overline{[Na^+]}$ | Nominal plasma sodium concentration [mEq/l] | I | 142 | 142 [84] |
| $\bar{J}_{RR,ADH}$ | Nominal water reabsorption rate in the collecting ducts [·] | SS | 0.955 (0.01) | 0.97 [64] |
| $\bar{Q}_{PD}$ | Protein influx post burn [g/h] | S | 85.8 | - |
| $\lambda_{PD}$ | Protein influx decay rate [1/h] | S | 10 | - |
| $M_{P_{BT}}$ | Maximum burnt tissue hydrostatic pressure perturbation [mmHg] | S | 56 | 20-270 [70] |
| $\lambda_{1,P_{BT}}$ | Burnt tissue hydrostatic pressure perturbation slow decay rate [1/h] | S | 6.88 | - |
| $\mu$ | The ratio between slow decay rate to fast decay rate [·] | S | 365 | - |
| $\lambda_{2,P_{BT}}$ | Burnt tissue hydrostatic pressure perturbation fast decay rate [1/h] | I | $8\lambda_{1,P_{BT}}$ | - |
| $K_{1,EV}$ | Nominal tissue evaporation rate [ml/h·m$^2$] | I | 18.48 | 18.48 [72] |
| $\lambda_{1,EV}$ | Evaporation growth rate [1/h] | I | 0.073 | 0.073 [72] |
| $K_{2,EV}$ | Maximum evaporation rate [ml/h·m$^2$] | I | 28.68 | 28.68 [72] |
| $\lambda_{2,EV}$ | Evaporation decay rate [1/h] | I | -0.0052 | -0.0052 [72] |
| $\varepsilon_B$ | Fraction of body surface subject to burn [·] | S | 0.16-1 | dataset |
| $K_{EX}$ | Maximum exudation rate [ml/h·m$^2$] | I | 25 | 25 [71] |
| $\lambda_{EX}$ | Exudation decay rate [1/h] | I | -0.0038 | -0.0038 [71] |
| $\eta_{EX}$ | Exudate to tissue albumin ratio [·] | S | 0.60 | 0.75 [21] |
| $\bar{\alpha}$ | Nominal albumin to capillary pore radius ratio [·] | SS | 0.83 (0.9) | 0.7-0.9 [32] |
| $k_{PD,BT}$ | Capillary destruction fraction for burnt tissue [·] | S | 0.53 | 0.50 [19] |
| $k_{PD,IT}$ | Capillary destruction fraction for intact tissue [·] | I | 1 | - |
| $M_{\alpha_{BT}}$ | Maximum pore ratio perturbation in burnt tissue [·] | SS | 0.27 (0.08) | 0.30 [28] |
| $M_{\alpha_{IT}}$ | Maximum pore ratio perturbation in intact tissue [·] | SS | 0.21 (0.08) | 0.19 [85] |
| $\lambda_{1,\alpha}$ | Pore ratio slow decay rate [1/h] | S | 0.015 | 0.025 [21] |
| $\lambda_{2,\alpha}$ | Pore ratio fast decay rate [1/h] | I | $\mu\lambda_{1,\alpha}$ | - |
| $M_{P_C}$ | Maximum capillary hydrostatic pressure perturbation [mmHg] | SS | 18.62 (17) | 23 (5) [73] |
| $\lambda_{1,P_C}$ | Capillary hydrostatic pressure perturbation slow decay rate [1/h] | SS | 0.51 (0.33) | - |
| $\lambda_{2,P_C}$ | Capillary hydrostatic pressure perturbation fast decay rate [1/h] | I | $\mu\lambda_{1,P_C}$ | - |

# Acknowledgment

This research was supported in part by the U.S. Army SBIR Program [Award No. W81XWH-16-C-0179], the Congressionally Directed Medical Research Programs [Award No. W81XWH-19-1-0322], and the U.S. National Science Foundation CAREER Award [Award No. 1748762].

# References

[1] "Burn Incidence Fact Sheet – American Burn Association." [Online]. Available: https://ameriburn.org/who-we-are/media/burn-incidence-fact-sheet/. [Accessed: 29-Nov-2020].

[2] C. B. Nielson, N. C. Duethman, J. M. Howard, M. Moncure, and J. G. Wood, "Burns: Pathophysiology of Systemic Complications and Current Management," *J. Burn Care Res.*, vol. 38, no. 1, pp. e469–e481, Jan. 2017.

[3] F. N. Williams *et al.*, "The leading causes of death after burn injury in a single pediatric burn center.," *Crit. Care*, vol. 13, no. 6, p. R183, 2009.

[4] "American Burn Association, National Burn Repository® 2016. Version 12.0., Chicago, IL."

[5] K. D. Capek *et al.*, "Contemporary Burn Survival," *J. Am. Coll. Surg.*, vol. 226, no. 4, pp. 453–463, Apr. 2018.

[6] T. J. Schaefer and O. Nunez Lopez, *Burns, Resuscitation And Management*. StatPearls Publishing, 2019.

[7] D. G. Greenhalgh, "Burn resuscitation: The results of the ISBI/ABA survey," *Burns*, vol. 36, no. 2, pp. 176–182, Mar. 2010.

[8] R. Cartotto, D. G. Greenhalgh, and C. Cancio, "Burn state of the science: Fluid resuscitation," *J. Burn Care Res.*, vol. 38, no. 3, pp. e596–e604, 2017.

[9] L. C. Cancio, J. Salinas, and G. C. Kramer, "Protocolized Resuscitation of Burn Patients," *Critical Care Clinics*, vol. 32, no. 4. W.B. Saunders, pp. 599–610, 01-Oct-2016.

[10] J. Salinas *et al.*, "Computerized decision support system improves fluid resuscitation following severe burns: An original study," *Crit. Care Med.*, vol. 39, no. 9, pp. 2031–2038, 2011.

[11] R. Cartotto and A. Zhou, "Fluid Creep: The Pendulum Hasn't Swung Back Yet!," *J. Burn Care Res.*, vol. 31, no. 4, pp. 551–558, Jul. 2010.

[12] M. B. Klein *et al.*, "The association between fluid administration and outcome following major burn: A multicenter study," *Ann. Surg.*, vol. 245, no. 4, pp. 622–628, Apr. 2007.

[13] R. G. Carlson, R. K. Finley, S. F. Miller, L. M. Jones, M. A. Morath, and S. Alkire, "Fluid retention during the first 48 hours as an indicator of burn survival," *J. Trauma - Inj. Infect. Crit. Care*, vol. 26, no. 9, pp. 840–844, 1986.

[14] B. Parvinian *et al.*, "Credibility evidence for computational patient models used in the development of physiological closed-loop controlled devices for critical care medicine," *Front. Physiol.*, vol. 10, no. MAR, p. 220, Mar. 2019.

[15] T. M. Morrison, P. Pathmanathan, M. Adwan, and E. Margerrison, "Advancing Regulatory Science With Computational Modeling for Medical Devices at the FDA's Office of Science and Engineering Laboratories," *Front. Med.*, vol. 5, no. SEP, p. 241, Sep. 2018.

[16] P. Glynn, S. D. Unudurthi, and T. J. Hund, "Mathematical modeling of physiological systems: An essential tool for discovery," *Life Sciences*, vol. 111, no. 1. Elsevier Inc., pp. 1–5, 2014.

[17] J. L. Bert, B. D. Bowen, R. K. Reed, and R. K. R. Microvascular, "Microvascular exchange and interstitial volume regulation in the rat: model validation."

[18] J. L. Bert, B. D. Bowen, X. Gu, T. Lund, and R. K. Reed, "Microvascular exchange during burn injury: II. Formulation and validation of a mathematical model.," *Circ. Shock*, vol. 28, no. 3, pp. 199–219, Jul. 1989.

[19] J. L. Bert, B. D. Bowen, R. K. Reed, and H. Onarheim, "Microvascular exchange during burn injury: IV. Fluid resuscitation model.," *Circ. Shock*, vol. 34, no. 3, pp. 285–97, Jul. 1991.

[20] S. L. Xie, R. K. Reed, B. D. Bowen, and J. L. Bert, "A Model of Human Microvascular Exchange," *Microvascular Research*, vol. 49, no. 2. pp. 141–162, 1995.

[21] R. T. Ampratwum, B. D. Bowen, T. Lund, R. K. Reed, and J. L. Bert, "A model of fluid resuscitation following burn injury: formulation and parameter estimation," *Comput. Methods Programs Biomed.*, vol. 47, no. 1, pp. 1–19, 1995.

[22] G. Arturson, T. Groth, A. Hedlund, and B. Zaar, "Potential use of computer simulation in treatment of burns with special regard to oedema formation," *Scand. J. Plast. Reconstr. Surg. Hand Surg.*, vol. 18, no. 1, pp. 39–48, 1984.

[23] A. Hedlund, B. Zaar, T. Groth, and G. Arturson, "Computer simulation of fluid resuscitation in trauma. I. Description of an extensive pathophysiological model and its first validation," *Comput. Methods Programs Biomed.*, vol. 27, no. 1, pp. 7–21, 1988.

[24] G. Arturson, T. Groth, A. Hedlundy, and B. Zaar, "Computer simulation of fluid resuscitation in trauma: First pragmatic validation in thermal injury," *J. Burn Care Rehabil.*, vol. 10, no. 4, pp. 292–299, 1989.

[25] L. M. Roa, T. Gomez-Cia, and A. Cantero, "Analysis of burn injury by digital simulation," *Burns*, vol. 14, no. 3, pp. 201–209,

The final version of this paper has been accepted for publication in the IEEE journal of Transactions on Biomedical Engineering and is available for early access at https://ieeexplore.ieee.org/document/9478222

DOI: 10.1109/TBME.2021.30945151988.

[26]  L. Roa Romero and T. Gomez Cia, "Analysis of the extracellular protein and fluid shifts in burned patients," *Burns*, vol. 12, no. 5, pp. 337–342, 1986.

[27]  L. Roa and T. Gómez-Cia, "A burn patient resuscitation therapy designed by computer simulation (BET). Part 1: simulation studies," *Burns*, vol. 19, no. 4, pp. 324–331, 1993.

[28]  T. Lund, H. Onarheim, and R. K. Reed, "Pathogenesis of edema formation in burn injuries," *World J. Surg.*, vol. 16, no. 1, pp. 2–9, 1992.

[29]  Q. Huang, M. Zhao, and K. Zhao, "Alteration of vascular permeability in burn injury," *Med. Express*, vol. 1, no. 2, pp. 62–76, 2014.

[30]  H. Wiig and M. A. Swartz, "Interstitial fluid and lymph formation and transport: Physiological regulation and roles in inflammation and cancer," *Physiological Reviews*, vol. 92, no. 3. pp. 1005–1060, 01-Jul-2012.

[31]  G. Arabidarrehdor *et al.*, "Mathematical model of volume kinetics and renal function after burn injury and resuscitation," *Burns*, Jul. 2020.

[32]  C. Chapple, B. D. Bowen, R. K. Reed, S. L. Xie, and J. L. Bert, "A model of human microvascular exchange: parameter estimation based on normals and nephrotics," *Comput. Methods Programs Biomed.*, vol. 41, no. 1, pp. 33–54, 1993.

[33]  J. Salinas, M. Serio-Melvin, C. Fenrich, K. Chung, G. Kramer, and L. Cancio, "225," *Crit. Care Med.*, vol. 40, pp. 1–328, Dec. 2012.

[34]  D. L. McGee, "Weight-height relationships and body mass index: Some observations from the diverse populations collaboration," *Am. J. Phys. Anthropol.*, vol. 128, no. 1, pp. 220–229, Sep. 2005.

[35]  B. K. Kataria *et al.*, "The Pharmacokinetics of Propofol in Children using Three Different Data Analysis Approaches," *Anesthesiology*, vol. 80, pp. 104–122, 1994.

[36]  A. Tivay, X. Jin, A. Lo, C. G. Scully, and J.-O. Hahn, "Practical Use of Regularization in Individualizing a Mathematical Model of Cardiovascular Hemodynamics Using Scarce Data," *Front. Physiol.*, 2020.

[37]  M. J. Muller, S. P. Pegg, and M. R. Rule, "Determinants of death following burn injury," *Br. J. Surg.*, vol. 88, no. 4, pp. 583–587, 2001.

[38]  R. L. George *et al.*, "The association between sex and mortality among burn patients as modified by age," *J. Burn Care Rehabil.*, vol. 26, no. 5, pp. 416–421, 2005.

[39]  G. McGwin, R. L. George, J. M. Cross, D. A. Reiff, I. H. Chaudry, and L. W. Rue, "Gender differences in mortality following burn injury," *Shock*, vol. 18, no. 4, pp. 311–315, 2002.

[40]  G. E. O'Keefe, J. L. Hunt, and G. F. Purdue, "An evaluation of risk factors for mortality after burn trauma and the identification of gender-dependent differences in outcomes," *J. Am. Coll. Surg.*, vol. 192, no. 2, pp. 153–160, 2001.

[41]  S. Meshulam-Derazon, S. Nachumovsky, D. Ad-El, J. Sulkes, and D. J. Hauben, "Prediction of morbidity and mortality on admission to a burn unit," *Plast. Reconstr. Surg.*, vol. 118, no. 1, pp. 116–120, Jul. 2006.

[42]  R. E. Barrow, M. G. Jeschke, and D. N. Herndon, "Early fluid resuscitation improves outcomes in severely burned children," *Resuscitation*, vol. 45, no. 2, pp. 91–96, 2000.

[43]  V. K. Tiwari, "Burn wound: How it differs from other wounds," *Indian Journal of Plastic Surgery*, vol. 45, no. 2. pp. 364–373, May-2012.

[44]  P. Wurzer, D. Culnan, L. C. Cancio, and G. C. Kramer, "Pathophysiology of burn shock and burn edema," in *Total Burn Care: Fifth Edition*, Elsevier Inc., 2018, pp. 66-76.e3.

[45]  R. H. Demling, "The Burn Edema Process : Current Concepts," *J. Burn Care Rehabil.*, vol. 26, pp. 207–227, 2005.

[46]  I. Kaddoura, G. Abu-Sittah, A. Ibrahim, R. Karamanoukian, and N. Papazian, "Burn injury: review of pathophysiology and therapeutic modalities in major burns.," *Ann. Burns Fire Disasters*, vol. 30, no. 2, pp. 95–102, Jun. 2017.

[47]  B. A. Latenser, "Critical care of the burn patient: The first 48 hours," *Crit. Care Med.*, vol. 37, no. 10, pp. 2819–2826, 2009.

[48] M. Lehnhardt *et al.*, "A qualitative and quantitative analysis of protein loss in human burn wounds," *Burns*, vol. 31, no. 2, pp. 159–167, 2005.

[49] H. J. Zdolsek, B. Kågedal, B. Lisander, and R. G. Hahn, "Glomerular filtration rate is increased in burn patients," *Burns*, vol. 36, no. 8, pp. 1271–1276, Dec. 2010.

[50] M. E. Habib *et al.*, "Does Ringer Lactate Used in Parkland Formula for Burn Resuscitation Adequately Restore Body Electrolytes and Proteins?," *Mod. Plast. Surg.*, vol. 07, no. 01, pp. 1–12, Jan. 2017.

[51] G. D. Warden, "Fluid resuscitation and early management," in *Total Burn Care*, 4th ed., Elsevier (2012), pp. 115–124.

[52] A. J. Diver, "The evolution of burn fluid resuscitation," *International Journal of Surgery*, vol. 6, no. 4. Elsevier, pp. 345–350, 01-Aug-2008.

[53] B. H. BOWSER-WALLACE, J. B. CONE, and F. T. CALDWELL, "Hypertonic Lactated Saline Resuscitation of Severely Burned Patients Over 60 Years of Age," *J. Trauma Inj. Infect. Crit. Care*, vol. 25, no. 1, pp. 22–26, Jan. 1985.

[54] N. T. Dai *et al.*, "The comparison of early fluid therapy in extensive flame burns between inhalation and noninhalation injuries," *Burns*, vol. 24, no. 7, pp. 671–675, 1998.

[55] M. Rani and M. G. Schwacha, "Aging and the pathogenic response to burn," *Aging Dis.*, vol. 3, no. 2, pp. 171–180, 2012.

[56] R. Oakley and B. Tharakan, "Vascular hyperpermeability and aging," *Aging Dis.*, vol. 5, no. 2, pp. 114–125, 2014.

[57] J. A. Farina, M. J. Rosique, and R. G. Rosique, "Curbing inflammation in burn patients," *Int. J. Inflam.*, vol. 2013, 2013.

[58] R. H. Demling, "Smoke inhalation lung injury: an update.," *Eplasty*, vol. 8, p. e27, May 2008.

[59] S. Rehou, S. Shahrokhi, J. Thai, M. Stanojcic, and M. G. Jeschke, "Acute Phase Response in Critically Ill Elderly Burn Patients," *Crit. Care Med.*, vol. 47, no. 2, pp. 201–209, Feb. 2019.

[60] M. S. Gregory, D. E. Faunce, L. A. Duffner, and E. J. Kovacs, "Gender difference in cell-mediated immunity after thermal injury is mediated, in part, by elevated levels of interleukin-6," *J. Leukoc. Biol.*, vol. 67, no. 3, pp. 319–326, 2000.

[61] E. H. Bresler and L. J. Groome, "On equations for combined convective and diffusive transport of neutral solute across porous membranes.," *Am. J. Physiol.*, vol. 241, no. 5, pp. F469-76, Nov. 1981.

[62] A. H. Øien and H. Wiig, "Modeling In Vivo Interstitial Hydration-Pressure Relationships in Skin and Skeletal Muscle," *Biophys. J.*, vol. 115, no. 5, pp. 924–935, 2018.

[63] Guyton and Hall, "The Urinary System: Functional Anatomy and Urine Formation by the Kidneys," in *Medical Physiology*, .

[64] F. M. Toates and K. Oatley, "Computer simulation of thirst and water balance," *Med. Biol. Eng.*, vol. 8, no. 1, pp. 71–87, 1970.

[65] F. M. Toates and K. Oatley, "Control of water-excretion by antidiuretic hormone: Some aspects of modelling the system," *Med. Biol. Eng. Comput.*, vol. 15, no. 6, pp. 579–588, 1977.

[66] Guyton and Hall, "Urine Formation by the Kidneys: II. Tubular Reabsorption and Secretion," in *Textbook of Medical Physiology*, .

[67] P. J. G. M. Voets and R. P. P. W. M. Maas, "Extracellular volume depletion and resultant hypotonic hyponatremia: A novel translational approach," *Math. Biosci.*, vol. 295, no. May 2017, pp. 62–66, 2018.

[68] F.-R. Curry, C. Michel, E. Renkin, and S. Geiger, "Mechanics and thermodynamics of transcapillary exchange." 01-Jan-1984.

[69] M. Jarzynska and M. Pietruszka, "The application of the Kedem-Katchalsky equations to membrane transport of ethyl alcohol and glucose," *Desalination*, vol. 280, no. 1–3, pp. 14–19, 2011.

[70] T. Lund, H. Onarheim, H. Wiig, and R. K. Reed, "Mechanisms behind increased dermal imbibition pressure in acute burn edema," *Am. J. Physiol. - Hear. Circ. Physiol.*, vol. 256, no. 4, 1989.

[71] J. Baudoin, P. Jafari, J. Meuli, L. A. Applegate, and W. Raffoul, "Topical negative pressure on burns: An innovative method for wound exudate collection," *Plast. Reconstr. Surg. - Glob. Open*, vol. 4, no. 11, 2016.

[72] R. X. Xu, X. Sun, and B. S. Weeks, *Burns regenerative medicine and therapy*. Karger, 2004.

[73] R. M. Pitt, J. C. Parker, G. J. Jurkovich, A. E. Taylor, and P. W. Curreri, "Analysis of altered capillary pressure and permeability after thermal injury," *J. Surg. Res.*, vol. 42, no. 6, pp. 693–702, 1987.

[74] J. P. Coghlan, J. S. Fan, B. A. Scoggins, and A. A. Shulkes, "Measurement of extracellular fluid volume and blood volume in sheep.," *Aust. J. Biol. Sci.*, vol. 30, no. 1–2, pp. 71–84, Apr. 1977.

[75] J. D. Hoppe, P. C. Scriba, and H. Klüter, "Transfusion Medicine and Hemotherapy - Chapter 5. Human Albumin," in *Transfusion medicine and hemotherapy : offizielles Organ der Deutschen Gesellschaft fur Transfusionsmedizin und Immunhamatologie*, vol. 36, no. 6, Karger Publishers, 2009, pp. 399–407.

[76] M. Ellmerer *et al.*, "Measurement of interstitial albumin in human skeletal muscle and adipose tissue by open-flow microperfusion," *Am. J. Physiol. - Endocrinol. Metab.*, vol. 278, no. 2 41-2, 2000.

[77] A. C. Guyton, A. E. Taylor, and H. J. Granger, "Dynamics and control of the body fluids, chapter 9-11," in *Circulatory physiology*, no. 2, Saunders, 1975, pp. vii, 397 p.

[78] J. Hall and A. Guyton, "Cardiac Failure," in *Textbook of Medical Physiology*, p. 276.

[79] J. Hall and A. Guyton, "Urine Formation by the Kidneys: I. Glomerular Filtration, Renal Blood Flow, and Their Control," in *Textbook of Medical Physiology*, .

[80] M. Nesje, A. Flåøyen, and L. Moe, "Estimation of glomerular filtration rate in normal sheep by the disappearance of iohexol from serum," *Vet. Res. Commun.*, vol. 21, no. 1, pp. 29–35, 1997.

[81] G. Baumann and J. F. Dingman, "Distribution, blood transport, and degradation of antidiuretic hormone in man," *J. Clin. Invest.*, vol. 57, no. 5, pp. 1109–1116, 1976.

[82] H. HELLER and S. M. ZAIDI, "The metabolism of exogenous and endogenous antidiuretic hormone in the kidney and liver in vivo.," *Br. J. Pharmacol. Chemother.*, vol. 12, no. 3, pp. 284–292, 1957.

[83] A. C. Gordon and J. A. Russell, "Should Vasopressin Be Used in Septic Shock?," in *Evidence-Based Practice of Critical Care*, Elsevier Inc., 2011, pp. 212–217.

[84] Guyton and Hall, "The Body Fluid Compartments: Extracellular and Intracellular Fluids; Edem," in *Medical Physiology*, .

[85] L. Kongstad, A. D. Möller, and P. O. Grände, "Reflection coefficient for albumin and capillary fluid permeability in cat calf muscle after traumatic injury," *Acta Physiol. Scand.*, vol. 165, no. 4, pp. 369–377, Apr. 1999.